\documentclass[12pt]{iopart}

\usepackage{epsfig}

\begin{document}
\title[Critical properties of the contact process with quenched dilution]
{Critical properties of the contact process with quenched dilution}

\author{Alexander H. O. Wada and M\'{a}rio J. de Oliveira}

\address{Instituto de F\'{\i}sica,
Universidade de S\~{a}o Paulo, \\
Rua do Mat\~ao, 1371,
05508-090 S\~{a}o Paulo, S\~{a}o Paulo, Brazil}

\ead{oliveira@if.usp.br}

\begin{abstract}

We have studied the critical properties of the contact process 
on a square lattice with
quenched site dilution by Monte Carlo simulations.
This was achieved by generating in advance the percolating cluster,
through the use of an appropriate epidemic model, and then by
the simulation of the contact process on the top of the percolating cluster.
The dynamic critical exponents were calculated by assuming
an activated scaling relation and the static exponents 
by the usual power law behavior. 
Our results are in agreement with the prediction that the quenched diluted
contact process belongs to the universality class of the random
transverse-field Ising model.
We have also analyzed the model and determined the phase diagram
by the use of a mean-field theory that takes into account the
correlation between neighboring sites. 

\end{abstract}

\maketitle

%-------------------------------------------------------------
\section{Introduction}

In many experiments on condensed matter,
quenched disorder may be present either because it is an
unavoidable feature of the sample or because disorder
is deliberated introduced in the sample. In either case, if we wish to
describe the properties of these systems by statistical mechanical
models, quenched disorder should be taken into account in these models.
In some cases the quenched disorder is irrelevant in 
the sense that it does not change the critical behavior.
In other cases, the quenched disorder is a relevant feature
that changes the critical behavior of the pure system.
According to a criterion due to Harris \cite{abharris1974}, 
a spatially quenched disorder will be irrelevant with respect
to the critical properties if the inequality $d\nu_\perp>2$ 
is obeyed for the pure system, where $\nu_\perp$  is the spatial 
correlation length exponent and $d$ is the dimension of the system.
For models belonging to the directed percolation universality class,
such as the contact process \cite{teharris1974,marro1999,henkel2008,tome2015},
this inequality is not fulfilled for $d<4$. We should thus expect
a change in the critical properties of the 
contact process with quenched disorder, as is the case of the
quenched diluted contact process,
which is the object of our study here.
Numerical simulations of the quenched diluted contact process
in two dimensional lattices
\cite{moreira1996,dickman1998,vojta2005,dahmen2007,oliveira2008,vojta2009}
indeed confirm the change in the critical properties.

A remarkable critical behavior of the 
quenched diluted contact process is the slow 
activated dynamics, of the logarithmic type, instead of the usual power law type. 
This result was advanced by Hooyberghs et al. \cite{hooyberghs2003,hooyberghs2004}
by mapping the evolution operator of the stochastic process 
describing the quenched diluted contact process into a random quantum spin-1/2
operator, and by the use of a renormalization group approach.
This critical behavior places the quenched diluted contact process into
the universality class of the random transverse-field Ising model
\cite{fisher1992,fisher1999,motrunich2000,lin2000,karevski2001,hoyos2008,
kovacs2010,miyasaki2013}.
The slow activated dynamics of the quenched diluted contact process
has been confirmed by numerical simulations in two dimensions 
\cite{vojta2005,dahmen2007,oliveira2008,vojta2009}.

Here we study a quenched diluted contact process in which the
quenched dilution is obtained by the
removal of a fraction of sites of the lattice. The remaining sites
form then clusters of site percolation. 
We aim to study the critical properties of the
contact process with quenched dilution by a method in which
the percolation clusters are understood as related to the stationary state
of stochastic models for the spreading of disease \cite{tome2010,tome2011,wada2015}.
The use of an epidemic process turns out to be a
procedure to create percolation clusters as efficient as the ordinary
method of simply creating random vacancies and then using a
clustering algorithm to find the percolating cluster.  
%new material
A straightforward numerical approach to the diluted contact process
is to consider all the remaining sites of a lattice after a certain 
fraction of them has been removed
\cite{moreira1996,dickman1998,oliveira2008}. 
Other methods such as ours consider instead just the sites
of the percolating cluster \cite{dahmen2007,vojta2009}.
In this case the total computer time should include the 
time it takes to generate the percolating cluster.
However, this time is very short, representing 
in our approach less than 1\% of the total computer time.

The stochastic model we use to generate clusters of site percolation
is defined as follows \cite{tome2011,wada2015}.
Each site of a regular lattice is occupied
by an individual that can be in one of three states:
susceptible, exposed or immune.
A susceptible individual, in the presence of an exposed individual, 
becomes exposed with certain probability $p$ and immune with the complementary
probability. The exposed and immune individuals remain forever
in these states. Starting with a single exposed individual in a lattice
full of susceptible individuals, a cluster of exposed individual
is generated such that at the stationary state it is exactly mapped
into a cluster of site percolation \cite{tome2011,wada2015},
with $p$ being identified as the probability of a site occupation.

Once a cluster of site percolation is generated by the model of 
spreading of disease explained above, we simulate the contact
process in the top of the percolating cluster. Only the percolating cluster
is needed because a finite cluster cannot sustain an active state.
That is, if we wait enough time, the absorbing state will be reached.
This procedure is thus interpreted
as the contact process with quenched site dilution. 
More details on the models will be given in the next section.
In this same section we set up the evolution
equations for one and two-site correlations and solve them by
the use a pair mean-field approximation, which allows us to construct
the phase diagram. This phase diagram shows that at the
percolation threshold the critical creation rate of the 
quenched diluted contact process is finite. 

Using the method presented above, we have obtained the critical
properties and the phase diagram of the diluted contact process
by numerical simulations and also by a mean-field theory. 
The method allowed us to obtain more accurate values for
the critical exponents and thus confirming the prediction that
the quenched diluted contact process belongs to the universality
class of the random transverse-field Ising model.
The mapping of the epidemic processes into the quenched dilution
contact process, allows also to conclude that this universality 
class may include some models for epidemic spreading.

%new material
The contact process, and other models belonging to the universality
of directed percolation, describes the transition of an active
state to an absorbing state, in which a system cannot never scape.
This nonequilibrium phase transition is very common in nature
and may occur in various situations \cite{hinrichsen2000}.
However, the experimental observation of the critical exponents
is very difficult, as any amount of disorder should alter the
critical behavior, but the exponents were eventually measured
in a electrohydrodynamic convection of nematic liquid crystals
\cite{takeuchi2007}. 

%-------------------------------------------------------------
\section{Models and pair approximation}

\begin{figure}
\centering
\epsfig{file=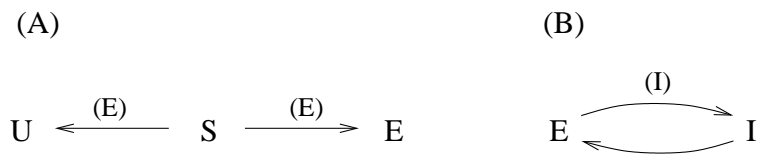,width=10cm}
\caption{
(A) The processes S $\to$ U and S $\to$ E are catalytic 
and represent the reactions S+E $\to$ U+E and S+E $\to$ E+E,
respectively. (B) The process E $\to$ I is catalytic 
and represents the reactions S+E $\to$ U+E. The reaction I $\to$ E is
spontaneous.}
\label{ilust}
\end{figure}

We begin by defining the two models by using the spreading of 
disease language. The two models are illustrated in figure
\ref{ilust}. The first model (A) is the one that generates the 
site percolation clusters, and is thus the underlying support
over which the second model (B), the contact process, is defined. 

\subsection{First model}

Each site of a regular lattice is occupied by an individual
that can be susceptible (S), immune (U) or exposed (E).
The possible processes of the first model are as follows:
\begin{equation}
{\rm S + E} \to {\rm U + E}, \qquad {\rm rate} \quad a,
\end{equation}
\begin{equation}
{\rm S + E} \to {\rm E + E}, \qquad {\rm rate} \quad b.
\end{equation}
These two processes define a continuous time stochastic process
whose probability distribution obeys a master equation.
Instead of writing down the master equation, which gives the
time evolution of the probability distribution, we
write the time evolution of some marginal probability
distribution such as the one-site and two-site probability
distributions. Using a procedure developed earlier \cite{wada2015,tome2011}
and the notations $P_X$, $P_{XY}$, $P_{XYZ}$ for one-site,
two-site and three-site probabilities, the following
time evolution equations can be derived
\begin{equation}
\frac{d}{dt}P_S = -(a+b) P_{ES},
\end{equation}
\begin{equation}
\frac{d}{dt}P_E = b P_{ES},
\end{equation}
\begin{equation}
\frac{d}{dt}P_{ES} = -\frac{a+b}{k} P_{ES} -(a+b)\mu P_{ESE} + b \mu P_{ESS},
\end{equation}
\begin{equation}
\frac{d}{dt}P_{US} = -(a+b) \mu P_{ESU} + a \mu P_{ESS},
\end{equation}
\begin{equation}
\frac{d}{dt}P_{EU} = \frac{a}{k} P_{ES} + a\mu P_{ESE} + b \mu P_{ESU},
\end{equation}
where $k$ is the coordination number of the regular 
lattice and $\mu=(k-1)/k$. 

%-----------------------------------
%---------------------------------

An approximate solution can be obtained by the use of
the pair mean-field approach which amounts to use the
approximation $P_{XYZ}=P_{XY}P_{YZ}/P_Y$. Using
the notation $x=P_S$, $y=P_E$, $v=P_{ES}$, $u=P_{US}$, 
and $w=P_{EU}$, we may write
\begin{equation}
\frac{dx}{dt} = -(a+b) v,
\label{5a}
\end{equation}
\begin{equation}
\frac{dy}{dt}= b v,
\label{5b}
\end{equation}
\begin{equation}
\frac{dv}{dt} = -\frac{a+b}{k} v -(a+b)\mu \frac{v^2}{x}
+ b \mu \frac{v(x-v-u)}{x},
\label{5c}
\end{equation}
\begin{equation}
\frac{du}{dt} = -(a+b)\mu \frac{vu}{x} + a \mu \frac{v(x-v-u)}{x},
\label{5d}
\end{equation}
\begin{equation}
\frac{dw}{dt} = \frac{a}{k} v + a\mu \frac{v^2}{x} + b \mu \frac{vu}{x},
\label{5e}
\end{equation}
where we have taken into account that $P_{SS}=P_S-P_{ES}-P_{US}=x-v-u$.
Equations (\ref{5a})-(\ref{5e}) have been solved in reference 
\cite{tome2011}. At the stationary state, the solution is 
\begin{equation}
x = s^k,
\label{7a}
\end{equation}
\begin{equation}
y = p(1-s^k),
\label{7b}
\end{equation}
\begin{equation}
v = 0,
\label{7c}
\end{equation}
\begin{equation}
u = qs^{k-1}(1-s^{k-1}),
\label{7d}
\end{equation}
\begin{equation}
w = pq(1-s^{k-1}),
\label{7e}
\end{equation}
where $s$ is the root of the polynomial equation
\begin{equation}
p \,s^{k-1} - s + q = 0,
\label{17}
\end{equation}
and $p=b/(a+b)$ and $q=1-p$.
The trivial solution is $s=1$, which gives $x=1$, $y=v=u=w=0$ 
and corresponds to the non spreading regime.
The solution $s\neq1$ corresponds to the
spreading regime and occurs only when $p>p_c=1/(k-1)$.
We remark that the stationary solution is exactly mapped
into the site percolation model with $p$ playing the role
of the probability of occupancy or the fraction of occupied sites
in the percolation model. The spreading regime ($s\neq1$) corresponds to
the existence of the percolating cluster. The non-spreading regime
corresponds to the absence of the percolating cluster ($s=1$).

\subsection{Second model}

As before, an individual can be susceptible (S), immune (U)
or exposed (E). In addition,  an individual can
also be infected (I). Thus in the second model, each
site can be in one of the states: S, U, E, and I.
However, the sites in states S and U remains forever in these states.
The only sites that have their states modified are the E and I sites.
They are modified according to the following processes 
\begin{equation}
{\rm E + I} \to {\rm I + I}, \qquad {\rm rate} \quad c,
\label{15a}
\end{equation}
\begin{equation}
{\rm I} \to {\rm E}, \qquad\qquad\quad\, {\rm rate} \quad r,
\label{15b}
\end{equation}
which are the reactions of the contact process. The relation between 
the infection rate $\lambda$, often used in studies
of the contact process, and $c$ and $r$ is given by $\lambda=c/r$.
For convenience, we also make use of a parameter $\alpha$, defined
by $\alpha=r/c=\lambda^{-1}$.

The initial state of the second model is chosen to be the stationary state
of the first model.
However, this state has no site in state I and the dynamics does not
start. To start the dynamics we choose randomly one site in state E
and replace it by a state I. By this procedure, a cluster of sites in
state I growths in the top of the percolation cluster. It should be
understood that the percolation cluster is formed by sites of type E and I.
The sites of type U are at the border of the percolation cluster.
The rest of the sites are in state S, and they are not connected to
the sites of the percolation cluster.

The two reactions (\ref{15a}) and (\ref{15b}) show
that the number of sites E and sites I is a constant
implying that the sum $P_E+P_I$ is a constant. Since these two
reactions do not involve the sites U and S, it follows
that the number of sites U and the number of sites S
are invariants, and the sum $P_{EU}+P_{IU}$ is a constant.

Again, using the procedure developed earlier \cite{wada2015,tome2011},
the following time evolution equations for the one-site and two-site
probabilities can be obtained
\begin{equation}
\frac{d}{dt} P_I = - r P_I + c P_{IE},
\end{equation}
\begin{equation}
\frac{d}{dt}P_{IE} = - r P_{IE} + r P_{II} - \frac{c}{k}P_{IE}
- c\mu P_{IEI}  + c\mu P_{IEE},
\end{equation}
\begin{equation}
\frac{d}{dt}P_{IU} = - r P_{IU} + c\mu P_{IEU},
\end{equation}
where $\mu=(k-1)/k$.
Due to the constraints stated above it is not necessary to
write down the time evolution equations for the other
one-site and two-site probabilities.

%----------------------------
\begin{figure}
\centering
\epsfig{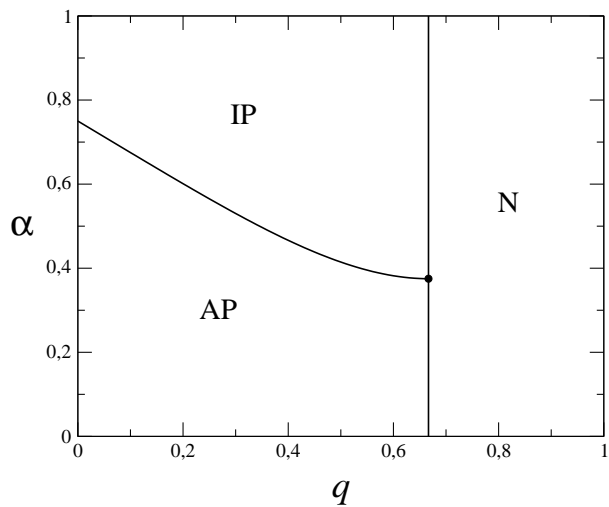}
\caption{Phase diagram from the pair approximation in
the plane $\alpha=\lambda^{-1}$ versus $q=1-p$, for a lattice of coordination $k=4$.
The phases are: active percolating (AP), inactive percolating (IP), and
nonpercolating (N).}
\label{aqpar}
\end{figure}
%---------------------------

Using again the pair approximation and the previous notation
together with the notations $z=P_I$, $g=P_{IE}$, $h=P_{IU}$, we may write
\begin{equation}
\frac{dz}{dt} = - r z + c g,
\label{20a}
\end{equation}
\begin{equation}
\frac{dg}{dt} = - r g + r (z-g-h) - \frac{c}{k} g
- c\mu \frac{g^2}{y}  + c\mu \frac{g}{y}(y-g-w),
\label{20b}
\end{equation}
\begin{equation}
\frac{dh}{dt} = - r h + c\mu \frac{gw}{y},
\label{20c}
\end{equation}
where we have taken into account that
$P_{II}=P_I-P_{IE}-P_{IU}=z-g-h$ and
$P_{EE}=P_E-P_{EI}-P_{EU}=y-g-w$.

Equations (\ref{20a})-(\ref{20c}) are to be solved using as 
initial conditions the stationary state of the 
first model. Since $P_E+P_I=y+z$ is invariant, it follows that $y+z=y_0$
where $y_0$ is the value of $P_E$ at the stationary state
of the first model, given by equation (\ref{7b}). 
Analogously, $P_{EU}+P_{IU}=w+h$ is
invariant implying $w+h=w_0$ where $w_0$ is the value of $P_{EU}$
at the stationary state of the first model,
given by equation (\ref{7e}).
At stationary state, equations (\ref{20a})-(\ref{20c})
have a trivial solution $z=0$, characterizing the absorbing state,
and a nontrivial solution for which $z\neq0$, characterizing the
active state. Solving for $z$, it is possible to obtain 
an expression for the nontrivial solution $z$. By taking
the limit $z\to0$ of the nontrivial solution we get the
critical line, which is given by
\begin{equation}
\alpha = \frac{r}{c} = \frac{k-1}{k}\left(1-\frac{q(1-s)}{p(1-s^k)}\right),
\end{equation}
where $s$ is the root of the polynomial equation given by equation (\ref{17}).
The critical line $\alpha$ versus $q$, shown in figure \ref{aqpar},
separates the active percolating phase from the inactive percolation phase.
Notice that, when $p\to p_c=1/(k-1)$ we get $\alpha=2(k-1)/k^2=\alpha_0$, so 
that the critical line meet the vertical line $p=p_c$ at $\alpha=\alpha_0$,
as shown in figure \ref{aqpar}.
It straightforward to show that
the critical exponent related to the order parameter is the
same as that of the pure system. 

\section{Scaling relations and numerical simulations}

Around the critical point, the quantities that
characterize the critical behavior are assumed to obey scaling relations.
In the present case of the diluted contact model, for which the
quenched disorder is relevant, the usual scaling relation
in terms of power laws in time is replaced by power laws
in the logarithm of time, called activated scaling \cite{vojta2009,hooyberghs2003}.
At the critical point, the space correlation length
$\xi$ behaves as \cite{hooyberghs2003}
\begin{equation}
\xi \sim (\ln t/t_0)^{1/\psi},
\label{26}
\end{equation}
where $\psi$ is the tunneling critical exponent \cite{fisher1999}
and $t_0$ is a constant.
Other quantities behave similarly at the critical point,
such as $N_I$, the number of infected individuals,
\begin{equation}
N_I \sim (\ln t/t_0)^{\theta},
\label{24}
\end{equation}
and $P$, the survival probability at time $t$,
\begin{equation}
P \sim (\ln t/t_0)^{-\delta}.
\label{25}
\end{equation}

From the scaling relations (\ref{26}), (\ref{24}),
and (\ref{25}) we find that
\begin{equation}
N_I \sim P^{-\theta/\delta},
\label{24a}
\end{equation}
\begin{equation}
N_I \sim \xi^{\,\theta\,\psi},
\label{25a}
\end{equation}
\begin{equation}
P \sim \xi^{-\delta\psi},
\label{26a}
\end{equation}
valid at the critical point.
These are useful relations because they do not depend
on $t_0$. 

%----------------------------
\begin{figure}
\centering
\epsfig{file=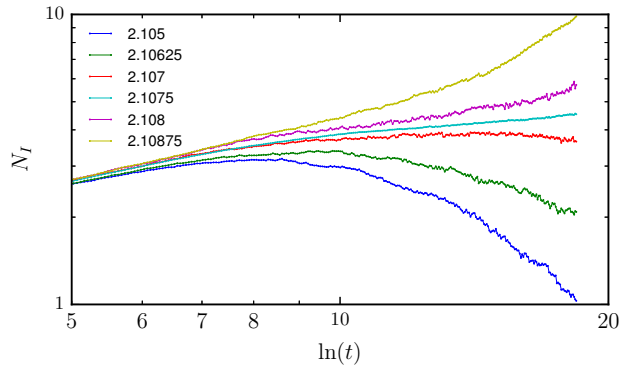,width=8.7cm}
\caption{Log-log plot of the number of infected sites $N_I$
versus $t$ obtained from numerical simulations at $p=0.8$
for the values of $\lambda$ indicated.}
\label{NIxt}
\end{figure}
%---------------------------

At the stationary state ($t\to\infty$), the quantities 
that describe the critical behavior follow the usual
power laws, but with exponents distinct from those of the pure system.
The order parameter $\rho$, defined as the fraction of infected sites
in the percolating cluster, behaves as 
\begin{equation}
\rho \sim (\lambda-\lambda_c)^\beta.
\label{29}
\end{equation}

Initially, we have simulated the first model to generate a percolating cluster.
The simulation, with periodic boundary conditions, was performed as follows.
At each time step, we choose at
random a bond from a list of active bonds. An active bond is a pair of 
SE nearest neighbor sites. The site S of the chosen bond becomes
E with probability $p$ and becomes U with the complementary probability
$q=1-p$. The chosen bond is removed from the list and the list us updated.
The time is then incremented by an amount $1/N_a$, where $N_a$ is the
number of active bonds in the list. Notice that, if a site S has $n_E$ nearest
neighbor sites in states E, then it will appear $n_E$ times in the list.
Starting with just one E site in a lattice full of S sites, this algorithm
will generate a cluster of E sites. This process stops when there is no
SE bonds in the lattice. When this happens the cluster of E sites is a site
percolating cluster with U sites standing in the border of the cluster,
separating the E sites from the S sites.

Having generated a percolating cluster of E sites, we simulate the 
contact model on top of the cluster using the following algorithm.
At each time step, a site is chosen at random among a list of
I sites. With probability $p_a=\lambda/(\lambda+1)$ it becomes an E site and with the
complementary probability $1-p_a$ a nearest neighbor site is chosen
at random. If the chosen neighboring site is in state E, it becomes I,
otherwise, nothing happens.
The time is then incremented by an amount $1/N_I$, where $N_I$ is the number
of I sites. The initial condition is formed by the cluster of E sites
with one E site turned into an I site.
This I site is taken as the origin.

%----------------------------
\begin{figure}
\centering
\epsfig{file=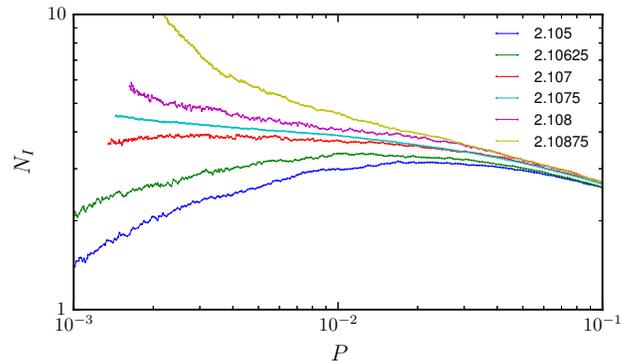,width=8.7cm}
\caption{Log-log plot of the number of infected sites $N_I$
as a function of the survival probability $P$ obtained from numerical
simulations at $p=0.8$ for the values of $\lambda$ indicated.}
\label{NIxP}
\end{figure}
%---------------------------
%----------------------------
\begin{figure}
\centering
\epsfig{file=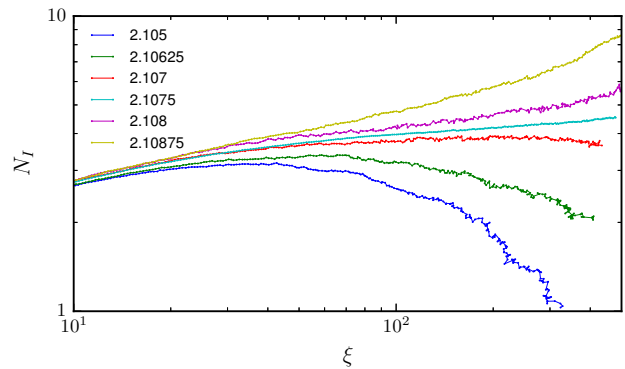,width=8.7cm}
\caption{Log-log plot of the number of infected sites $N_I$
as a function of the spatial correlation length $\xi$ obtained from numerical
simulations at $p=0.8$ for the values of $\lambda$ indicated.}
\label{NIxcsi}
\end{figure}
%---------------------------

We have performed simulations on a square lattice with $N=L^2$ sites
with $L$ up to $L=8192$. For several values $p$ and $\lambda$,
we have measured, as a function of time,
the number of infected sites $N_I$, the survival probability
$P$ and the correlation length $\xi$ defined by
\begin{equation}
\xi^2 = \frac1{N_I} \sum_i \langle r_i^2\rangle,
\end{equation}
where the summation is over the sites occupied by an infected
and $r_i$ is the distance from the site $i$ to the origin.
Each quantity was measured by averaging over $10^5$ to $10^6$ disorder
configurations, where each disorder configuration is obtained by
the simulation of the first model starting with a distinct seed of random number.
%new material
We have also performed simulation with smaller values of $L$. 
However, results coming from lattice with $L=4096$ agree,
within statistical errors and up to the maximum time 
we have used, with those coming from $L=8192$. The statistical
errors were determined by the calculation of the standard
statistical deviation. 
The results for the three quantities $N_I$, $P$ and $\xi$,
determined for several values of $\lambda$, are shown in
figures \ref{NIxt}, \ref{NIxP}, and \ref{NIxcsi}. 
The error bars in these figures are not shown, but they
are less than 8\%. At the critical point, $\lambda=2.1075$,
they are even less reaching 1\%.
%end of new material.

Figure \ref{NIxt} shows the plot of the number of infected $N_I$ as
a function of time $t$ for $p=0.8$.
Fitting the expression (\ref{24}) to the
data points of figure \ref{NIxt} we estimate the critical parameter as
being $\lambda=2.1075(1)$, the critical exponent as $\theta=0.13(2)$,
and $\ln t_0=6.0(5)$.
To find the exponents $\psi$ and $\delta$ and a better estimate of $\theta$
we use a procedure similar to the one used in \cite{vojta2009}
in which we first determine the quantities 
$\theta/\delta$, $\theta\,\psi$ and $\delta\psi$, by fitting the 
expressions (\ref{24a}), (\ref{25a}) and (\ref{26a}) to the data
points. After that, we use expressions (\ref{26}), (\ref{24}) and (\ref{25})
to find the exponents $\theta$, $\delta$ and $\psi$ by a constrained fitting,
to be explained below.

From the plots of $N_I$ versus $P$, shown in figure \ref{NIxP},
$N_I$ versus $\xi$, shown in figure \ref{NIxcsi}, 
and $P$ versus $\xi$,
we may get, respectively, $\theta/\delta$, $\theta\,\psi$ 
and $\delta\psi$. Since the scaling relations (\ref{24a})
(\ref{25a}) and (\ref{26a}) do not involve
time, the estimates of these quantities are independent
of the $t_0$, resulting in more precise values, which are found to be
\begin{equation}
\theta/\delta = 0.075(5),
%0.0746(52) 
\label{17a}
\end{equation}
\begin{equation}
\theta\,\psi = 0.078(4)
%0.0784(42), 
\label{17b}
\end{equation}
\begin{equation}
\delta\psi =  1.034(23).
%1.034(23)
\label{17c}
\end{equation}
The consistency of these values can be checked by dividing equations
(\ref{17b}) and (\ref{17c}). The result is 
$\theta\,\psi/\delta\psi = 0.0758(57)$, which is in fair agreement with (\ref{17a}). 
The value of $\theta/\delta$ can be used to get the ratio
$\beta/\nu_\perp$ between the order parameter critical exponents $\beta$
and the critical exponents $\nu_\perp$ related to the spatial correlation
length. Using the relation $\theta/\delta=d\nu_\perp/\beta-2$ we find
\begin{equation}
\beta/\nu_\perp=0.964(2).
%0.9640(24)
\end{equation}

%----------------------------
\begin{figure}
\centering
\epsfig{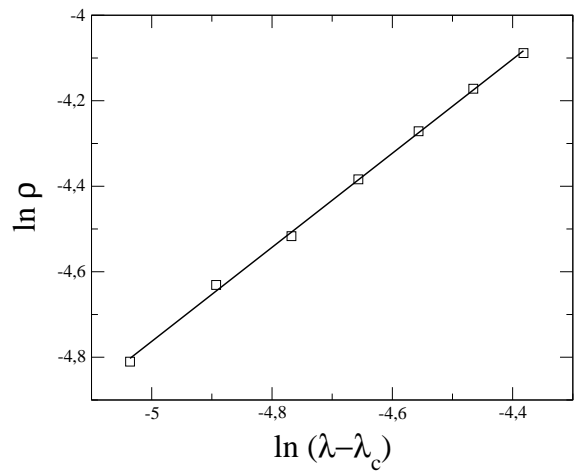}
\caption{Fraction of infected sites $\rho$ in the percolating cluster
as a function of $\lambda$. The slope of the strait line fitted to
the data points gives $\beta=1.11(6)$.}
\label{beta}
\end{figure}
%---------------------------
%----------------------------
\begin{figure}
\centering
\epsfig{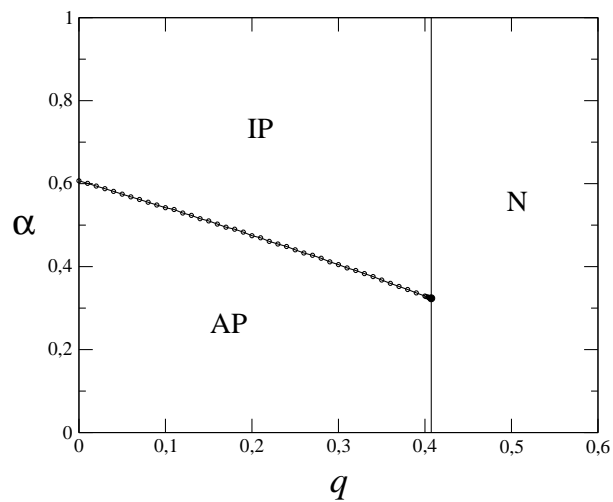}
\caption{Phase diagram from numerical simulations on a square lattice in
the plane $\alpha=\lambda^{-1}$ versus $q=1-p$.
The phases are: active percolating (AP), inactive percolating (IP), and
nonpercolating (N).}
\label{aqsim}
\end{figure}
%---------------------------

The exponents $\theta$, $\delta$ and $\psi$ are found by a
procedure as follows. For each value of $t_0$ in the
interval $5.5\leq t_0\leq 6.5$, we determine the exponents
$\theta$, $\delta$ and $\psi$ by fitting the expressions 
(\ref{26}), (\ref{24}) and (\ref{25}) to the data points.
After that we choose the actual values of these exponents
as the ones such that the quantities $\theta/\delta$, $\theta\,\psi$ 
and $\delta\psi$ are as close as possible
to the values given by (\ref{17a}), (\ref{17b}), (\ref{17c}).
This procedure leads to the following values for the exponents:
\begin{equation}
\theta = 0.145(8),
\end{equation}
\begin{equation}
\delta = 1.88(11),
\end{equation}
\begin{equation}
\psi = 0.55(3).
%\psi = 0.554(29).
\end{equation}

%------------ 
\begin{table}[t]
\begin{center} 
\caption{\label{exponents} Critical exponents obtained by
numerical simulations on a square lattice at $p=0.8$ and 
$\lambda_c=2.1075$ (second column). The third and fourth columns
show results for the quenched diluted contact process
\cite{oliveira2008,vojta2009} in $d=2$
whereas the last two columns show results for the 
random transverse-field Ising model 
\cite{motrunich2000,kovacs2010} also in $d=2$.
}
\bigskip
\begin{tabular}{|l|l|l|l|l|l|}
\hline
ref.      & this work &  \cite{oliveira2008} & \cite{vojta2009}
 & \cite{motrunich2000} & \cite{kovacs2010}\\
\hline
$\theta$         & $0.145(8)$&           &$0.15(3)$ &          &          \\
$\psi$           & $0.55(3)$ & $0.48(7)\,$ &$0.51(6)$ & $0.42(6)$&$0.48(2)$ \\
$\delta$         & $1.88(11)$  &           &$1.9(2)$  &          &          \\
$\phi$           & $1.87(10)$  &           &          & $2.5(4)$ &          \\
\hline
$\beta/\nu_\perp$& $0.964(2)$ & $0.95(2)\,$ &$0.96(2)$ & $1.0(1)$ &$0.982(15)$\\
$\beta$          & $1.11(6)$ &           &$1.15(9)$ &          &           \\
$\nu_\perp$      & $1.15(6)$ &           &$1.20(15)$&$1.07(15)$&$1.24(2)$  \\
$d_F$            & $1.036(2)$ &           &          & $1.0(1)$ &$1.018(15)$\\
\hline
\end{tabular}
\end{center}
\end{table}
%----------

We have also performed simulations to get the stationary properties
by using systems of linear size $L=2048$.
The interest quantities were obtained by the use
of $10^8$ Monte Carlo steps after discarding $10^7$ Monte Carlo steps.
In this case, we use as the initial state a configuration in which
a fraction of sites is in the infected state. Again we determined
the number of infected sites $N_I$ at the stationary state from which
we obtained the density $\rho=N_I/N_C$ where $N_C$ is the number of
sites of the cluster, that is, the number of I sites plus the number
of E sites. Assuming the critical behavior (\ref{29}),
we get the value $\beta=1.11(6)$ by plotting $\rho$ as a function
of $\lambda-\lambda_c$, as shown in figure \ref{beta} for the
case of $p=0.8$. This result for $\beta$ together with
the numerical value for the ratio $\beta/\nu_\perp$, obtained
above, gives us $\nu_\perp=1.15(6)$. 

We have also performed similar simulations for other value of $p$ and obtained
the critical line, shown in figure \ref{aqsim}. In particular, at the 
percolation critical point $p=p_c=0.59274$, we get $\lambda=3.10(1)$.
% $\alpha=0.322(1)$

The critical exponents obtained here are shown
in table \ref{exponents} together with results
coming from other papers on the quenched diluted contact process
\cite{vojta2009,oliveira2008}
and on the random transverse-field Ising model \cite{motrunich2000,kovacs2010}.
To make contact with exponents used to describe the critical
behavior of the random transverse-field Ising model,
we have determined from our results
the fractal dimension critical exponent $d_F=d-\beta/\nu_\perp$
and the exponent $\phi$, related to the fractal dimension and
the tunneling exponent by $d_F=\phi\psi$ \cite{motrunich2000}.
We see that our results agree, within the statistical errors,
to all other results cited in table \ref{exponents}.
The results are $d_F=1.036(2)$ and $\phi=1.87(10)$.
%$d_F=1.0360(24)$ and $\phi=1.87(10)$

%-------------------------------------------------------------
\section{Conclusion}

We have studied the critical properties of the
quenched diluted contact process through a mean-field
theory and Monte Carlo simulations by using a two stage procedure.
The first was the generation of the percolating cluster, obtained
by the use of a stochastic lattice model whose
stationary states are the clusters of percolation
model. The second stage was the simulation of the
contact process on the top of the percolating cluster.
It should be remarked that, only the percolating clusters
is necessary if we wish to study the static stationary properties
because finite clusters cannot support
an active state. For a finite cluster, the absorbing 
state will be reached if we wait enough time. As to the
dynamic properties, our results show that they can also
be obtained from the percolating cluster, or at least
their critical properties, as can be inferred by comparing
our critical exponents with other works. 
The present method allowed to obtain more precise critical
exponents, with errors that are at most equal to 6\%,  confirming 
the prediction that
the quenched diluted contact process belongs to the universality class of
the random transverse-field Ising model. 

\begin{figure}
\centering
\epsfig{file=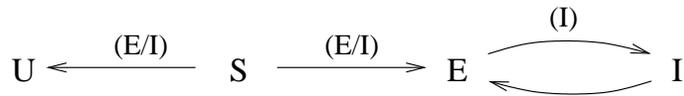,width=9cm}
\caption{
The processes S $\to$ U and S $\to$ E are catalytic 
with E or I playing the role of catalysts.
The process E $\to$ I is autocatalytic 
and the reaction I $\to$ E is spontaneous.}
\label{ilususei}
\end{figure}

The mapping of the two epidemic models into the quenched diluted
contact process allows to speculate about the existence of epidemic
models that are in the universality class of transverse-field Ising model.
In fact, this is the case of the model illustrated in figure \ref{ilususei},
which may be thought as a merger of the two models in figure \ref{ilust}.
The epidemic model of figure \ref{ilususei} is defined on a lattice
in which each site can be in one of four states: S, U, E, and I,
and is composed by three catalytic reactions: S $\to$ U, S $\to$ E,
E $\to$ I, and by a spontaneous reaction I $\to$ E.   
At the stationary states, the I and E sites form a connected cluster
of sites consisting of a site percolation cluster. The E and I sites
evolves then as the contact process on the top of a percolating cluster.
Therefore, the model defined by rules of figure \ref{ilususei}
is also mapped into the quenched diluted contact process and
its critical properties puts the model in the universality class
of the transverse-field Ising model.

%-----------------------------------
%---------------------------------

%%-------------------------------------------------------------
\section*{Acknowledgment}

We wish to acknowledge the Brazilian agency FAPESP for 
financial support.

%-------------------------------------------------------------


\section*{References}

\begin{thebibliography}{99}

\bibitem{abharris1974} A. B. Harris, J. Phys. C {\bf 7}, 1671 (1974).

\bibitem{teharris1974} T. E. Harris, Ann. Prob. {\bf 2}, 969 (1974).

\bibitem{marro1999} J. Marro and R. Dickman, {\it Noequilibrium
Phase Transitions in Lattice Models}, (Cambridge University Press,
Cambridge, 1999).

\bibitem{henkel2008} M. Henkel, H. Hinrichesen and S. L\"ubeck,
{\it Non-Equilibrium Phase Transitions}, Vol. I:
{\it Absorbing Phase Transitions} (Springer, Dordrecht, 2008).

\bibitem{tome2015} T. Tom\'e and M. J. de Oliveira, {\it Stochastic
Dynamics and Irreversibility}, Springer, 2015.

\bibitem{moreira1996} A. G. Moreira and R. Dickman, 
Phys. Rev. E {\bf 54}, R3090 (1996).

\bibitem{dickman1998} R. Dickman and A. G. Moreira, 
Phys. Rev. E {\bf 57}, 1263 (1998).

\bibitem{vojta2005} T. Vojta and M. Dickison, Phys. Rev. E {\bf 72},
036126 (2005).

\bibitem{dahmen2007} S. R. Dahmen, L. Sittler and H. Hinrichsen,
J. Stat. Mech. P01011 (2007).

\bibitem{oliveira2008} M. M. de Oliveira and S. C. Ferreira,
J. Stat. Mech. P11001 (2008).

\bibitem{vojta2009} T. Vojta, A. Farquhar and J. Mast,
%%''Infinite-randomness critical point in the two-dimensional disordered contact process.'' 
Phys. Rev. E {\bf 79}, 011111 (2009).

%----------------------------------------------------------

\bibitem{hooyberghs2003} J. Hooyberghs, F. Igl\'oi, and C. Vanderzande,
%''Strong Disorder Fixed Point in Absorbing-State Phase Transitions. ''
Phys. Rev. Lett. {\bf 90}, 100601 (2003).

\bibitem{hooyberghs2004} J. Hooyberghs, F. Igl\'oi, and C. Vanderzande,
Phys. Rev. E {\bf 69}, 066140 (2004).

\bibitem{fisher1992} D. S. Fisher, Phys. Rev. Lett. {\bf 69}, 534 (1992).

\bibitem{fisher1999} D. S. Fisher, Physica A {\bf 263}, 222 (1999).

\bibitem{motrunich2000} O. Motrunich, S. C. Mau, D. Huse, and D. S. Fisher,
%''Infinite-randomness quantum Ising critical fixed points,''
Phys. Rev. B {\bf 61}, 1160 (2000).

\bibitem{lin2000} Y.-C. Lin, N. Kawashima, F. Igl\'oi, and H. Rieger, 
%''Numerical Renormalization Group Study of Random Transverse Ising Models
%in One and Two Space Dimensions'',
Prog. Theor. Phys. Supplement {\bf 138 } 479 (2000).

\bibitem{karevski2001} D. Karevski, Y-C. Lin, H. Rieger, N.
Kawashima and F. Igl\'oi,
%''Random quantum magnets with broad disorder distribution'',
Eur. Phys. J. B {\bf 20}, 267 (2001).

\bibitem{hoyos2008} J. A. Hoyos, Phys. Rev. E {\bf 78}, 032101 (2008).

\bibitem{kovacs2010} I. A. Kov\'acs and F. Igl\'oi,
Phys. Rev. B {\bf 82}, 054437 (2010).

\bibitem{miyasaki2013} R. Miyazaki and H. Nishimori,
%''Real-space renormalization-group approach to the random transverse-field
%Ising model in finite dimensions,''
Phys. Rev. E {\bf 87}, 032154 (2013).

%----------------------------------------------------------

\bibitem{tome2010} T. Tom\'e and R. M. Ziff, Phys. Rev. E {\bf 82}, 051921 (2010).

\bibitem{tome2011} T. Tom\'e and  M. J. de Oliveira,
J. Phys. A {\bf 44}, 095005 (2011).

\bibitem{wada2015} A. H. O. Wada, T. Tom\'e and M. J. de Oliveira,
J. Stat. Mech. P04014 (2015).

%----------------------------------------------------------

\bibitem{hinrichsen2000} H. Hinrichsen,
%'On possible realizations of directed percolation', 
Braz. J. Phys. {\bf 30}, 69 (2000).

\bibitem{takeuchi2007} K. A. Takeuchi, M. Kuroda, H. Chat\'e, and M. Sato, 
%''Directe percolation criticality in turbulent liquid crystals'',
Phys. Rev. Lett. {\bf 99}, 234503 (2007).


\end{thebibliography}
\end{document}